\documentclass[12pt]{elsart3}
\usepackage[dvips]{graphicx}
\begin{document}
\begin{frontmatter}

\title{Development of Large area Gamma-ray Camera with GSO(Ce) Scintillator Arrays and PSPMTs}
\author[Kyoto]{H.Nishimura\corauthref{cor}},
\corauth[cor]{corresponding author}
\ead{nisimura@cr.scphys.kyoto-u.ac.jp}
\author[Kyoto]{K.Hattori},
\author[Kyoto]{S.Kabuki},
\author[Kyoto]{H.Kubo},
\author[Kyoto]{K.Miuchi},
\author[Waseda]{T.Nagayoshi},
\author[Kyoto]{Y.Okada},
\author[Kobe]{R.Orito},
\author[Kyoto]{H.Sekiya},
\author[Kyoto]{A.Takada},
\author[ICRR]{A.Takeda},
\author[Kyoto]{T.Tanimori}
and \author[Kyoto]{K.Ueno}

\address[Kyoto]{Department of Physics, Graduate School of Science, Kyoto University, Kitashirakawa-oiwake, Sakyo, Kyoto, 606-8502, Japan}
\address[Waseda]{Advanced Research Institute for Science and Engineering, Waseda University, 17 Kikui-cho, shinjyuku, Tokyo, 162-0044, Japan}
\address[Kobe]{Department of Physics, Graduate School of Science and Technology, Kobe University, 1-1 Rokkoudai, Nada, Kobe, 657-8501, Japan}
\address[ICRR]{Kamioka Observatory, ICRR, University of Tokyo, 456 Higasi-mozumi, Hida-shi, Gifu, 506-1205, Japan}

\begin{abstract}

 We have developed a position-sensitive scintillation camera with a large 
area absorber for use as 
an advanced Compton gamma-ray camera.
At first we tested GSO(Ce) crystals.
We compared light output from the GSO(Ce) crystals under various conditions: 
the method of surface polishing, 
the concentration of Ce, and co-doping Zr.
As a result, we chose the GSO(Ce) crystals doped with only 0.5 mol\% Ce,
and its surface polished by chemical etching as the scintillator of our camera.
We also made 
a 16$\times$16 cm$^2$ scintillation camera which consisted of 
9 position-sensitive PMTs
(PSPMTs Hamamatsu flat-panel H8500 ),
the each of which had 8$\times$8 anodes with a pitch of 6 mm 
and coupled to  8$\times$8 arrays 
of pixelated 6$\times$6$\times$13 mm$^3$ GSO(Ce) scintillators. 
For the readout system of the 576 anodes of the PMTs, 
we used chained resistors to reduce the number of readout channels down to 48 
to reduce power consumption. 
The camera has a position resolution of less than 6mm and 
a typical energy resolution of 10.5\% (FWHM) at 662 keV  
at each pixel in a large area of 16$\times$16 cm$^2$.

Furthermore we constructed a 16$\times$16 array of 
3$\times$3$\times$13 mm$^3$ pixelated GSO(Ce) scintillators, 
and glued it to a PMT H8500.
This camera had the position resolution of less than 3mm, 
over an area of 5$\times$5 cm$^2$, except for some of the edge pixels;
the energy resolution was typically 13\% (FWHM) at 662 keV. 
\end{abstract}
\end{frontmatter}
 
\section{Introduction}
We have developed an advanced Compton camera 
for gamma-ray astronomy 
in the range of 100 keV to 20 MeV \cite{takada}.
It needs a scintillation camera 
as a detector for Compton-scattered gamma rays with a
 good energy and position resolution and a large area
because the resolution and the efficiency for the scattered gamma rays  
contribute to the angular resolution 
and the efficiency of the advanced Compton camera.
In addition, 
radiation hardness and a high counting-rate performance of the scintillation 
camera are required.
For these reasons, 
we chose a GSO(Ce) (Gd$_2$SiO$_5$:Ce) crystal as a scintillator,
and the PSPMT H8500 (Flat Panel PMT produced by Hamamatsu\cite{hamamatsu}) 
as a position-sensitive photon device.

GSO(Ce) has advantages in astronomical use, 
such as having a higher-Z, 
faster decay time than NaI(Tl),
a higher light output than BGO,
greater radiation hardness and less radioactivation 
than most other known scintillators.
Furthermore, 
GSO(Ce) can be easily cut and polished, 
since it is nonhydroscopic.

The PSPMT H8500 was recently developed as a promising device 
for nuclear physics and medicine,
for example, PET and SPECT \cite{pani} \cite{Gim} \cite{Herber} \cite{houiken}.  
It has 8$\times$8 anodes with a 6 mm pitch 
and 12-stage metal channel dynodes.
The advantage of this PMT is  
that it has a much smaller dead space and a larger effective area
than that of the previous multi-anode PMTs. 
The effective area of this PMT is 49$\times$49 mm$^2$,
which is 89\% of the package size.

In this paper 
we report on the results of measurements with pixelated GSO(Ce) scintillators and 
the performance (energy resolution and position resolution) of 
the scintillation camera developed by us. 
    
\section{Measurements of Pixelated GSO(Ce) Scintillators}

There are different aspects 
that characterize the performance of a pixelated GSO(Ce) scintillator.
One is the pixel size. 
The width of 6 mm and the thickness of 13 mm 
were chosen for our scintillation camera 
in order to fit the pitch of the anodes of PSPMTs, 
and the radiation length.
One of the other important issues is the method of surface polishing.
There are two established methods of polishing:
one is chemical etching, 
and the other is mechanical polishing.
It was reported 
that there was 
little difference between the performances of 
pixelated scintillators polished with these methods \cite{hitachi2}.  
However, mechanical polishing is more expensive 
than chemical etching.  
Another important issue is 
the concentration of Ce as a scintillation activity impurity, 
and additional dopants.
The light-decay time becomes faster 
as the concentration of Ce increases, 
although
increasing the concentration of Ce decreases the optical transmittance of the crystal.
It was also reported that 
doping at 200 ppm of Zr to GSO(Ce) improves 
the optical transmittance of the crystal \cite{hitachi3}.
	   
\begin{figure}[t]
\includegraphics[width=0.5\textwidth]{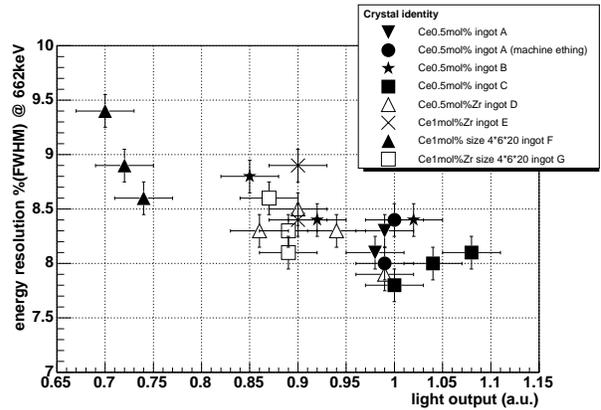}
\caption{Light output and energy resolution of pixelated GSO(Ce) scintillators with or without a Zr dopant. The systematic errors are also shown.}
\label{single}
\end{figure}

We measured the light outputs from several crystals 
under different conditions of polishing or doping impurity 
in order to examine 
how important these are. 
There were 8 types of pixelated scintillators,
which differed in the concentration of Ce, 
the presence of Zr 
and the ingot.
We enveloped each pixel scintillator by a reflector (Goatex)   
and coupled the crystal face with an area of
6$\times$6 mm$^2$ or 4$\times$6 mm$^2$  
to a single anode PMT (R6231 Hamamatsu) 
using optical grease (OKEN 6262).
We then irradiated it with 662 keV gamma rays 
from a $^{137}$Cs source through a $\phi$3 mm collimator.
Figure \ref{single} shows the relative light outputs 
and energy resolution 
at 662 keV.
The systematic errors were due to limits in the reproducibility of  
gluing the reflector to the crystal. 
It shows that the method of polishing 
and the optical transmittance caused by 
the concentration of Ce
are not more important for the performance of our pixel size of 6$\times$6$\times$13 mm$^3$ 
than the difference of the ingot. 
However,
there is a significant difference between only 1 mol\% Ce doped crystals 
with a size of 4$\times$6$\times$20 mm$^3$ and the others.  
This shows 
a significant decreasing of the optical transmittance caused by Ce
and the improvement of transmittance caused by doping Zr  
for longer crystal with a thickness of 20 mm.

Based on the above studies, 
we chose crystals that were polished by chemical-etching
and doped with only 0.5 mol\% Ce for our camera. 
We made an 8$\times$8 array of pixelated GSO(Ce) scintillators. 
Each pixel was optically separated by Vikuiti 3M ESR, 
which is a multilayer polymer mirror 
with a thickness of 65 $\mu$m and a reflectance of 98\%. 
The construction of this array is described in reference \cite{houiken}.
  

\section{Performance of a 5$\times$5 cm$^2$ scintillation camera and a 16$\times$16 cm$^2$ scintillation camera}
We made a 5$\times$5 cm$^2$ scintillation camera 
by coupling an 8$\times$8 GSO(Ce) array to a PMT H8500 using optical grease. 
In order to limit power consumption,
the readout circuit of the camera consisted of 8 resistive chains,
16 ch amplifiers, and ADCs, as described in reference \cite{sekiya}.

The image of each pixel scintillator was clearly resolved  
by a flood field of irradiation of 662 keV gamma ray,
which means that the position resolution was less than the pixel pitch of 6 mm.
We also obtained the energy spectrum of each pixel  
with energy resolution of 10\% (FWHM) @ 662 keV.
 
\begin{figure}[b]
\begin{center}
\includegraphics[width=0.4\textwidth, angle=0]{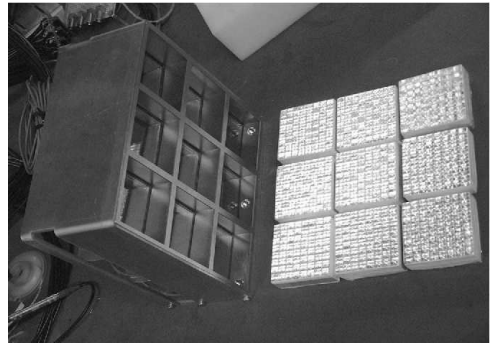}
\caption{Photograph of a 16$\times$16 cm$^2$ scintillation camera composed of 9 PSPMTs, the each of which coupled to 8$\times$8 arrays of pixelated 6$\times$6$\times$13 mm$^3$ GSO(Ce) crystals.}
\label{photo1}
\end{center}
\end{figure}

\begin{figure}[b]
\begin{center}
\includegraphics[width=0.45\textwidth]{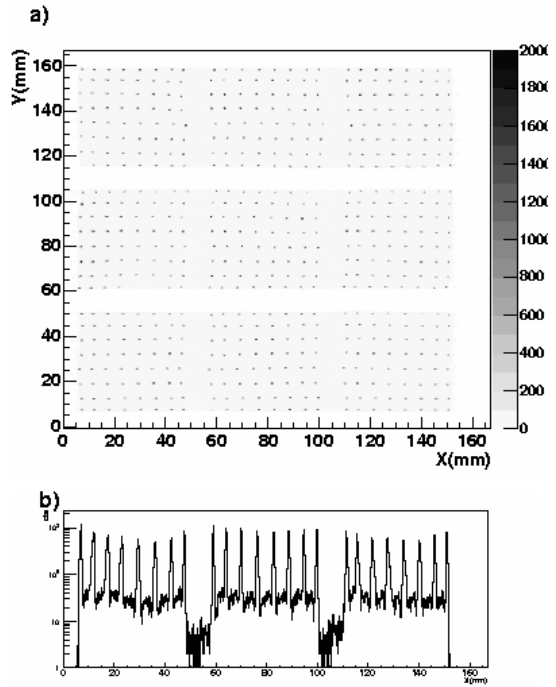}
\caption{(a)Event map measured with the 16$\times$16 cm$^2$ scintillation camera in a flood of irradiation of 662 keV gamma rays. (b)The logarithmic x-projection of the event map at the 12th row (78 mm$<$y$<$84 mm).}
\label{map_592}
\end{center}
\end{figure}

\begin{figure}[b]
\begin{center}
\includegraphics[width=0.45\textwidth, angle=0]{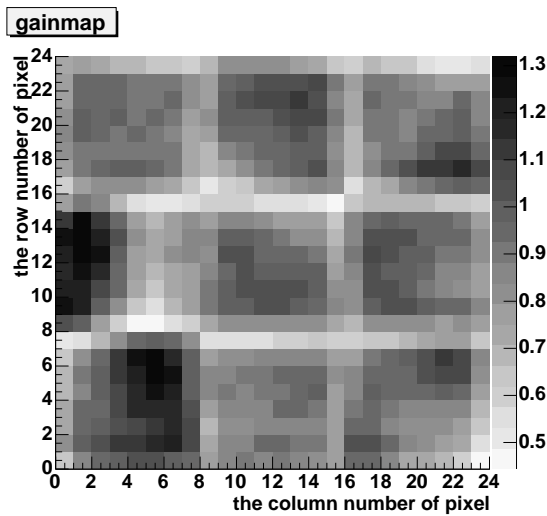}
\caption{Distribution of the light outputs for the 16$\times$16cm$^2$ scintillation camera composed of arrays of 6$\times$6$\times$13mm$^3$ GSO(Ce) pixelated crystals.}
\label{distribution}
\end{center}
\end{figure}

 
 This 5$\times$5 cm$^2$ camera can be easily extended to a larger camera.
We constructed a 16$\times$16 cm$^2$ camera with 3$\times$3 PMTs, 
as shown in Fig. \ref{photo1}.
The pitch of PMTs was 53 mm and the effective area of the camera is 82\%. 
The number of readout channels of the camera was 
only 48 channels with 24 resistive chains.
All 576 pixels
were clearly resolved by a flood field of radiation of 662 keV gamma rays, 
as shown in Fig. \ref{map_592} (a) and (b),
which show an event map and an x-projection map at the 12th row (78 mm$<$y$<$84 mm),
respectively. 
The events located between each pixels seem to be
multi-pixel hits events by Compton-scattered gamma rays or accidental events.
The energy resolution (FWHM) was 31.0\% at 122 keV, 
18.2\% @356 keV,
13.9\% @511 keV,
10.7\% @662 keV,
9.6\% @835 keV,
8.6\% @1173 keV for the typical pixel,
9.8\%@662 keV(FWHM) for good pixels 
and 13\%@662 keV(FWHM) for bad pixels.     
Figure \ref{distribution} shows 
a map of the relative light output of each pixel.
It mainly shows differences among anode gains of the PMTs.
However the light output was too low at some edges of each PMT. 
This was probably due to 
a misalignment of the array of crystals to the PMTs.   
The measurable energy ranges of this camera are 80-1300 keV and 100-900 keV,
for good and bad pixels, respectively.


\begin{figure}[htbp]
\includegraphics[width=0.4\textwidth, angle=0]{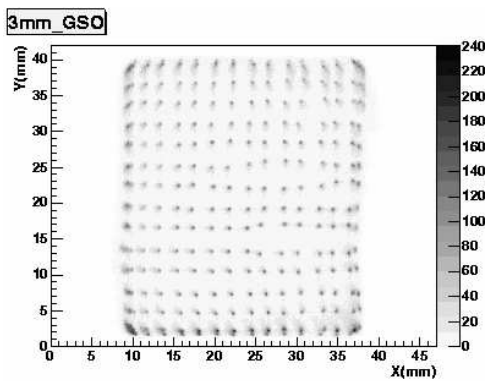}
\caption {Event map measured with a scintillation camera composed of a 16$\times$16 array of 3$\times$3$\times$13 mm$^3$ pixelated GSO(Ce) crystals and a H8500.}
\label{3mm_map}
\end{figure}

In order to improve the position resolution,
we tried to use smaller pixels with a width of 3mm 
compared to that of the anode pitch of H8500.
Such developments were 
already reported on earlier \cite{pani} \cite{Gim} \cite{Herber}.
We made a
16$\times$16 array of pixelated 
3$\times$3$\times$13 mm$^3$ GSO(Ce) scintillators 
and coupled it to the H8500.
In the flood field of irradiation of 662 keV gamma rays, 
the pixel image was clearly separated, 
except for some of edge pixels, 
as shown in Fig. \ref{3mm_map}.
However the energy resolution deteriorated to 12\% @662 keV   
compared to that
of scintillation cameras with 6$\times$6$\times$13mm$^3$ pixels.

\section {Summary}
We measured the light outputs from GSO(Ce) scintillators 
under different conditions,
and chose a non-Zr doped, chemical etched GSO(Ce) crystal 
with a size of 6$\times$6$\times$13 mm$^3$ for our scintillation camera.     
We constructed 8$\times$8 arrays of the crystal, 
and developed a 16$\times$16 cm$^2$ GSO scintillation camera.
The performance of this camera is 
sufficient to be used as a Compton-scattered gamma-ray camera
for our advanced Compton camera. 
We constructed an advanced Compton camera 
and tested its performance.    

\begin{ack}
This work is supported by a Grant-in-Aid for
the 21st Century COE ``Center for Diversity and
Universality in Physics''; 
a Grant-in-Aid in Scientific
Research of the Japan Ministry of Education,
Culture, Science, Sports, Technology;
``SENTAN'', JST.
We thank 
Dr. Y.Yanagida and
Hitachi Chemical Co., Ltd. 
for their support.
\end{ack}

\end{document}